# Study of Phonon Modes in Germanium Nanowires

**Xi Wang and Ali Shakouri [a)]**

*Baskin School of Engineering, University of California, Santa Cruz, CA 95064*

**Bin Yu, Xuhui Sun and Meyya Meyyappan**

*Center for Nanotechnology, NASA Ames Research Center Moffett Field, CA 94035*

**Abstract**

The observation of pure phonon confinement effect in germanium nanowires is limited due to the illumination sensitivity of Raman spectra. In this paper we measured Raman spectra for different size germanium nanowires with different excitation laser powers and wavelengths. By eliminating the local heating effect, the phonon confinement effect for small size nanowires was clearly identified. We have also fitted the Raman feature changes to estimate the size distribution of nanowires for the first time.

One-dimensional crystalline structures such as nanowires and nanotubes have been studied extensively during the past several years. Semi-conducting nanowires promise applications in

---

a) Author to whom correspondence should be addressed; electronic mail: ali@soe.ucsc.edu

future generation electronic and optoelectronic devices. Raman microscopy is a useful tool in the study of quasi-1D materials; it provides information about the surface and volume phonon modes and lattice vibrations, including how those vibrations are affected by extreme small dimensions. Recently, several papers [1, 9, 10] have analyzed the Raman peak shifts and the shape of the Raman spectrum for Si nanowires. However, the reported shifts and asymmetric broadenings vary depending on the experimental conditions. Studies show that the optical phonon peaks of Silicon nanowires are dependant on the excitation laser power and independent of wavelength. Thus, low laser power is essential in order to examine the phonon spectrum of different size nanowires[1].

Self-assembled single crystalline Germanium nanowires allow researchers to observe relatively strong one-dimensional confinement effects for both carriers and phonons. Compared to Si, Ge has smaller electron and hole effective masses and a lower dielectric constant; therefore, nanowires made of Ge should have stronger confinement characteristics than Si nanowires with the same diameters. The samples considered in this paper were synthesized on lithographically patterned Au catalyst arrays, with sizes ranging from 5~20 nm, by the Vapor-Liquid-Solid (VLS) method[2]. As a reference, a piece of bulk Germanium wafer was also examined under the same conditions. The ambient temperature was kept at typical room temperature, 22ºC. Fig. 1 shows the scanning electron microscopy (SEM) images of the as-synthesized Ge nanowires.

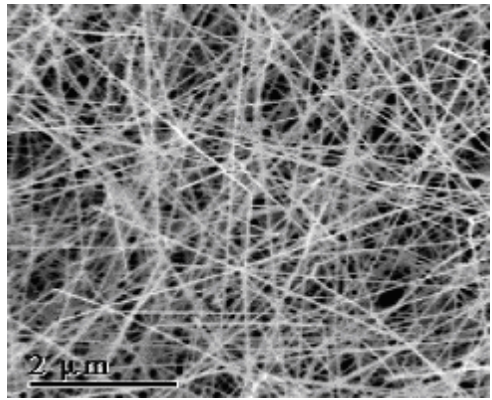

Fig. 1 A SEM image of the as-synthesized GeNW sample.

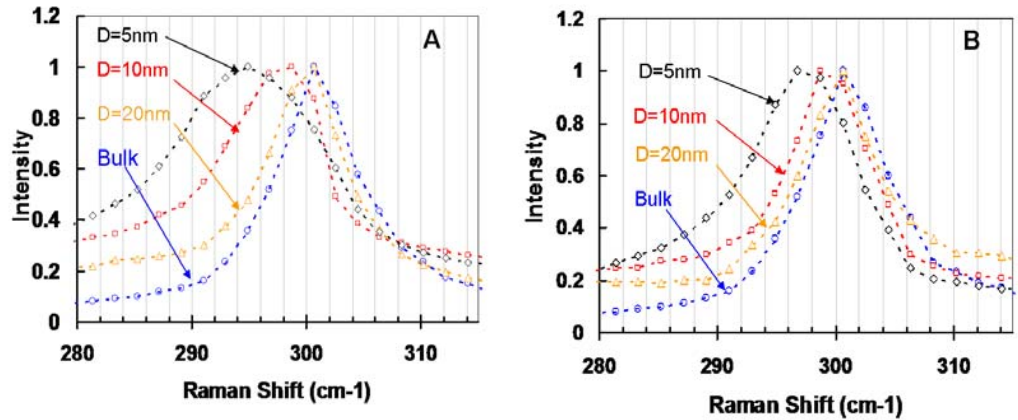

Fig. 2(A) Raman spectra of 3 GeNW samples (D=5, 10, 20nm) and bulk Ge measured at 500uW laser power with 514.523 nm wavelength. (B) Raman spectra of 3 GeNW samples (D=5, 10, 20nm) and bulk Ge measured at 50uW laser power with 514.523 nm wavelength.

Fig. 2(A) shows the evolution of Raman spectra as a function of catalyst size D. All four samples were excited by a 514.532nm, 500 $\mu W$, Ar$^+$ laser and the resulting scattered light was examined.. In comparing the spectrum from bulk Ge, we observed obvious position shift-downs, broadenings and increases of asymmetry from all Ge nanowire samples, from Stokes peaks in Ge range. As the wire size decreases, these features become more significant. Earlier papers[8,14~15] reported these phenomena as an entire contribution of scaling- induced phonon confinement effect. However, when we decreased the excitation laser power, we clearly observed much weaker Raman feature changes as a function of catalyst size. Raman spectra of the same four samples at 50uW excitation laser power level are shown in Fig. 2(B). We kept the ambient temperature stable at room temperature and the excitation times for each measurement equal. We also used a clean room for taking measurements in order to reduce contaminations. With all of these eliminations, excitation laser power difference became the only issue affecting the measured spectra. Pure confinement effect should be looked at only after carefully calibrating and removing this contribution.

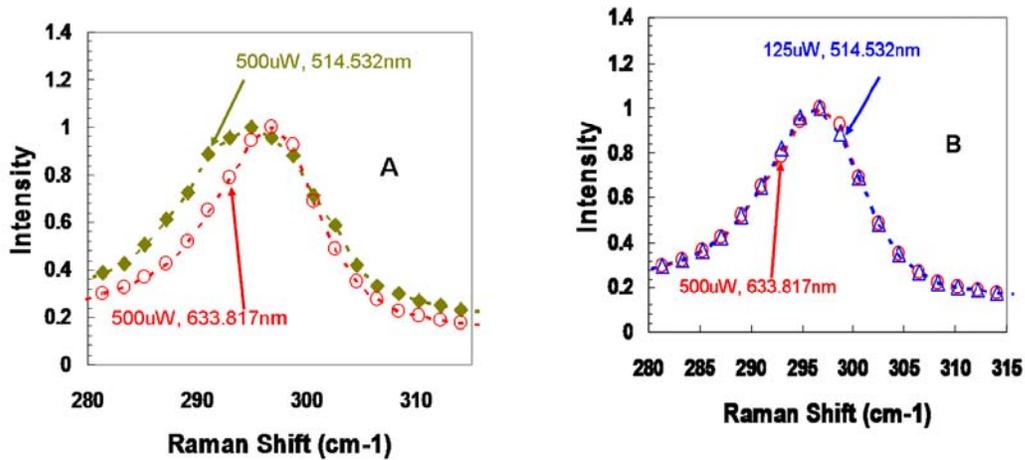

Fig. 3 (A) Comparison of Raman spectra of GeNW sample 3 (D=5nm) measured at 500μW with 514.532nm and 633.817nm excitation wavelength. (B) Comparison of Raman spectra of GeNW sample 3 (D=5nm) measured at 125μW with 514.532nm excitation wavelength and 500μW with 633.817nm excitation wavelength.

Two different excitation lasers, a 514.523 nm $Ar^+$ laser and a 633.817 nm $Kr^+$ laser, were used on the four samples to examine their wavelength independency. Fig. 3 (A-B) shows the results from GeNW sample 3 (D=5nm), which has the smallest dimension and thus is most sensitive to excitation condition changes. The first attempt involved exciting the sample with a laser having the same power level (500uW) but different excitation wavelength (514.523 and 633.817 nm). The results of this experiment do show an obvious Raman spectra change. However, it was hard to tell whether this change resulted from a difference in wavelength or a difference in the amount of heat absorbed at different wavelengths. Considering the fact that absorption coefficients of germanium are 600 $cm^{-1}$ for a 514.523nm $Ar^+$ laser and 150$cm^{-1}$ for a 633.817nm $Kr^+$ laser, we simply assume that, when applying the same excitation laser power, heat absorbed by Ge samples using a 633.817nm Kr+ laser is about ¼ that of the same sample using a 514.523nm $Ar^+$ laser. Fig3 (B) shows the Raman spectra of GeNW sample 3 (D=5nm) measured at 125μW with 514.532nm excitation wavelength and 500μW with 633.817nm excitation wavelength. The remarkable similarities of Raman features—position, FWHM, and asymmetry level— imply that the spectrum changes brought about by differing wavelengths are not due to resonant Raman selection of different size wires, but rather are the results of the absorbed power difference.

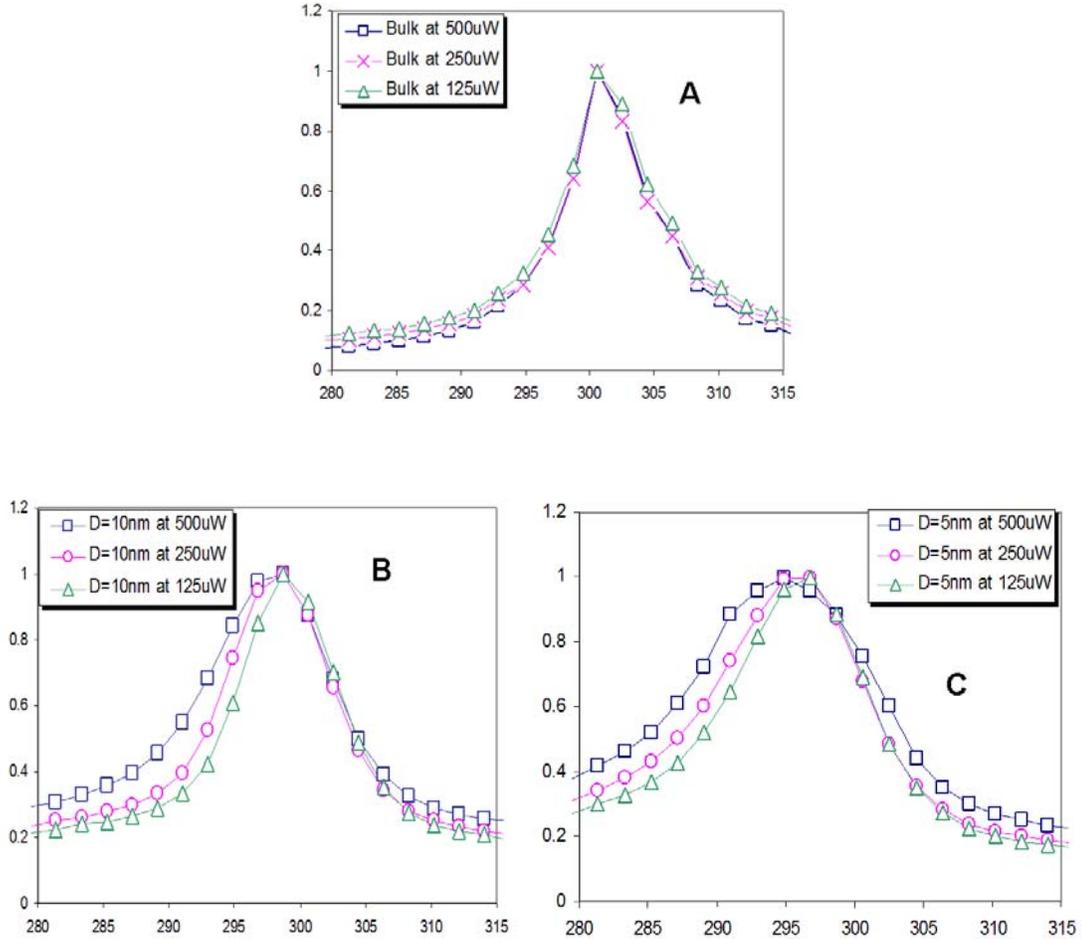

Fig. 4. (A) Raman Stokes peak of bulk Ge with different excitation powers; (B) Raman Stokes peak of 10nm GeNWs with different excitation powers; (C) Raman Stokes peak of 5nm GeNWs with different excitation powers;

To further understand how the excitation power interacts with nanostructure Raman spectra, each of the differently sized Ge nanowire samples was excited by three different laser powers (See Fig. 4 (A-C)). Neutral Density Filters (NSFs) were added between the sample and laser source to indicate the different excitation powers. From a $500\,\mu W$ source, the D0, D0.3 and D0.6 filters allow, respectively, $500\,\mu W$, $250\,\mu W$ and $125\,\mu W$ laser powers to pass through. As the power increases, the Raman Stokes peaks of Ge nanowires generally move to lower frequencies, broaden, and become less symmetric, while no significant change can be observed from those of the bulk Ge. This is consistent with what has been found for Si and Si nanowires[1]. Another interesting find is the difference between changing rates of Raman features from different sized catalyst samples. Smaller catalyst samples respond to excitation power change much more obviously than larger ones. This trend indicates excitation laser power affects nanostructure Raman features through localized heating, until this local heating disappears, as this is approximately what happens with the bulk material.

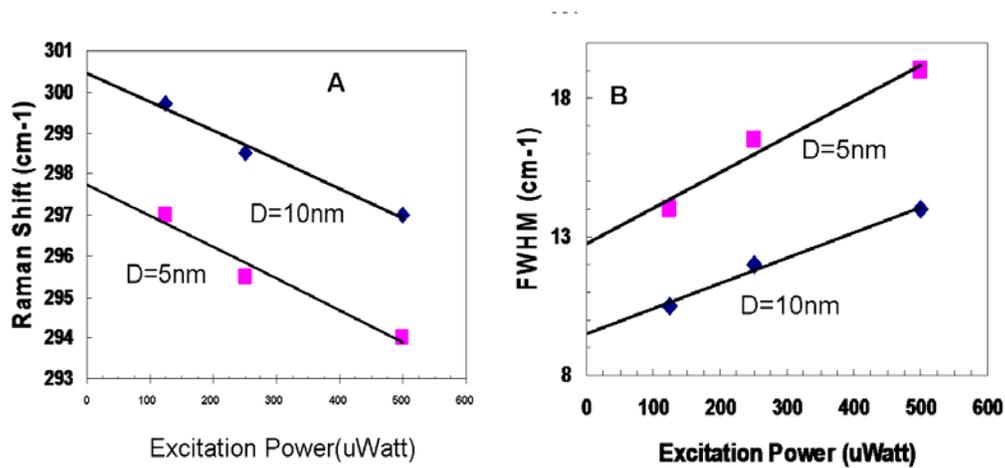

Fig.5 (A) Raman peak shift of GeNW samples with different excitation powers; (B) Raman peak FWHM of GeNW samples with different excitation powers;

Fig 5 shows the extracted Raman feature changes for 5 and 10 nm GeNW samples as a function of excitation laser power. With first order fitting, we can get the approximate Raman Stokes peak features for these GeNW at pure phonon confinement condition. The larger Raman feature changes from smaller wires reveals stronger phonon confinement effect in smaller wires; this theoretically will bring down the heat conduction capability of said wires. However, we need further evidence in order to prove this.

An estimation of sample local temperature can be obtained by monitoring the Stokes/anti-Stokes intensity ratio.

$$\frac{I_{AS}}{I_S} = \gamma \exp\left(\frac{k_B T}{h\nu}\right) \quad (1)$$

$T$ is the local temperature, $h\nu$ is the energy gap between excitation and ground states, and $\gamma$ is a coefficient relating to peak position and FWHM. Fig 6 shows the calculated Raman peak ratios as a function of excitation power for all samples. The Graph proves, firstly, that local heating induced by the highly focused laser beam at the excite point affects the measured spectra.. Secondly, the graph indicates that the heat accumulates more and creates a higher local temperature in smaller wires.

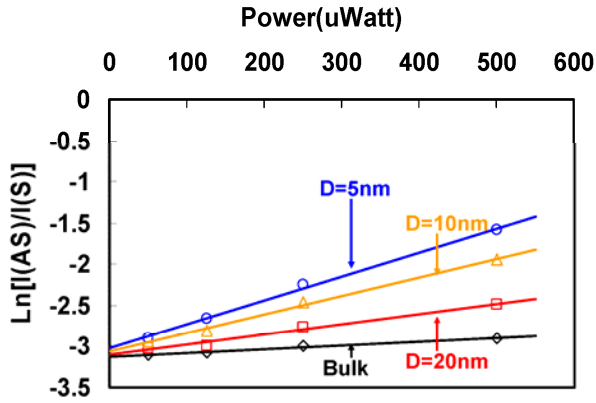

Fig .6 Raman peak ratio as a function of excitation laser power for 3 GeNW samples (D=5, 10, 20nm) and bulk Ge measured at 514.532nm wavelength.

Since the density of GeNW is approximately the same for each sample, the different local heating shall come from radius dependent thermal conductivity of nanowires.

$$\kappa = \frac{L Q_{cond}}{A \Delta T} \quad (2)$$

Equation (2) is the expression for thermal conductivity, where L represents the thickness of GeNW layer, $Q_{cond}$ is the amount of heat being conducted, A is the cross-section area through which the heat flows (laser spot size in our case), and $\Delta T$ is the local temperature change due to heat flow. From the relationship between Stokes/anti-Stokes intensity ratio and local temperature, we obtained

$$T \propto \ln(\frac{I_{AS}}{I_S}) \qquad (3)$$

Then,

$$\kappa \propto \frac{LQ_{cond}}{A\left[\ln(\frac{I_{AS}}{I_S})_1 - \ln(\frac{I_{AS}}{I_S})_2\right]} = qr \qquad (4)$$

$q = L/A$ is the dimension ratio from the heat path and $r = Q_{cond}/\left[\ln(I_{AS}/I_S)_1 - \ln(I_{AS}/I_S)_2\right]$ is the absolute slope of each line in Fig.6. Since we know that $\kappa = 59.9 W/m-K$ for intrinsic bulk germanium, we can estimate the GeNW layer thermal conductivity for each sample as $\kappa^* = \kappa_b r^*/r_b$. The results are 22.8, 12.1, and $9.1 W/m-K$ for 20, 10 and 5nm catalyst size GeNW, respectively. These results indicate a difference in heat conduction capabilities of different size nanowires. However, they should not be considered as actual thermal conductivities of GeNWs; there is, unfortunately, no way for us to decouple the heat absorbed by wires and the air in between.

Although the size of catalysts gives control to the size of synthesized nanowires, a variance still exists due to the nature of self-assembling growth. Therefore, it is good to confirm the actual nanowires sizes. Since it is hard to obtain distribution information from scanning microscopy, such as AFM, we used a fitting of Raman spectra. According to theoretical calculations based on the phonon confinement model of Richter *et al* and Campbell and Fauchet (RCF) [6, 7], the Raman intensity is given by

$$I(\omega) = \int \frac{|C(0,q)|^2}{[\omega - \omega(q)]^2 + \left(\frac{\Gamma_0}{2}\right)^2} d^3q \qquad (5)$$

where $C(0,q)$ is a Fourier coefficient of the confinement function, $\omega(q)$ is the phonon dispersion, q is the phonon momentum, and $\Gamma_0$ is FWHM of the reference Ge. Considering the basic shape and material of our nanowire samples, we used $|C(0,q)|^2 = \exp(-q^2D^2/16\pi^2)$, and $\omega(q) = [A+B\cos(qa/2)]^{0.5}+C$, with $A=0.69\times10^5 cm^{-2}$, $B=0.195\times10^5 cm^{-2}$. C is an adjustment parameter for reference samples measured in an experimental setting in this study, a is the lattice constant, and D is the interested diameter of GeNWs. Fig 7 shows the comparison between measurement and simulation based on RCF model. Similar results were obtained at D=20nm and D=100nm (bulk); however, slight differences were observed at D=5nm and D=10nm.

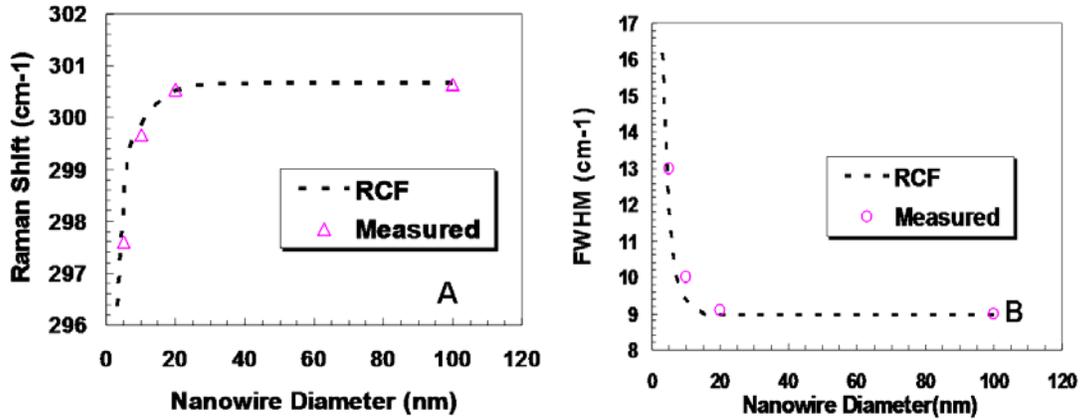

Fig. 7 (A) Comparison of Raman shift measured from 3 GeNW samples (D=5, 10, 20nm) and bulk Ge to calculation result form RCF model. (B) Comparison of Raman peak FWHM measured from 3 GeNW samples (D=5, 10, 20nm) and bulk Ge to calculation result form RCF model.

In order to extract the real nanowire diameter distribution, we fitted the measured spectra by using the RCF model. A least-square rule was applied as the fitting criterion. Fig 8 (A,B) show the fitting results: as-extracted nanowire diameter distributions are 18~19nm (for D=20nm sample), 7~9nm (for D=10nm sample) and 4~5nm (for D=5nm sample). The differences between catalyst sizes and actual nanowire diameters explain the differences between simulated and measured Raman features in Fig 8

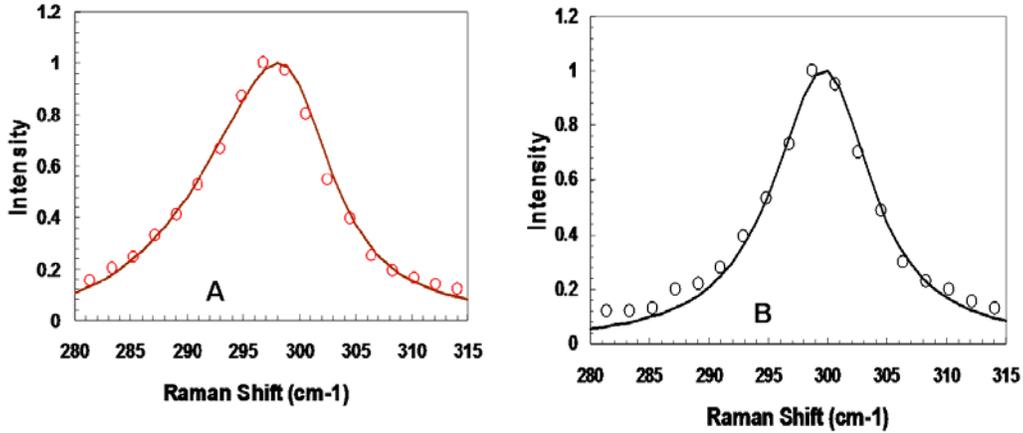

Fig.8 Dots: Raman spectra of GeNWs measured at 50μW from (A) Sample 3 (D=5nm) and (B) Sample 2 (D=10nm). Solid line: Fit to the measured spectra using RCF model adapted for GeNWs. Extracted diameter distributions are (A) 4~5nm and (B) 7~9nm.

In conclusion, we have shown the Raman spectra for different size germanium nanowires by measuring with differing laser powers and wavelengths. The study shows the excitation power dependency and wavelength independency of Raman spectrum evolution. By eliminating the heating of the sample under illumination, we can clearly identify the phonon confinement effect for small size nanowires. The fitting of Raman spectra was use to estimate the size distribution of nanowires for the first time. The results of GeNW layer conductivity calculation clearly indicate the different heat conduction capabilities of different sized nanowires.